\documentclass{article}
\usepackage{spconf,amsmath,graphicx}

\usepackage{enumitem}
\setlist{nosep, leftmargin=14pt}

\usepackage[noadjust]{cite}
\usepackage{amssymb,amsfonts}
\usepackage{algorithmic}
\usepackage{textcomp}
\usepackage{xcolor}
\usepackage{booktabs} 
\usepackage{subcaption}
\usepackage{multirow}
\usepackage{nicefrac}
\usepackage{pifont}
\usepackage[hidelinks]{hyperref}

\title{Retinal OCT Synthesis with Denoising Diffusion Probabilistic Models for Layer Segmentation\vspace{-0.1cm}}

\name{%
\begin{tabular}{cccc}
Yuli Wu$^{1}$$^{\star}$ & 
Weidong He$^{1,2}$$^{\star}$ & 
Dennis Eschweiler$^{1}$& 
Ningxin Dou$^{3}$ \\
Zixin Fan$^{3}$ &
Shengli Mi$^{2}$ &
Peter Walter$^{4}$ & 
Johannes Stegmaier$^{1}$\vspace{-0.3cm}
\end{tabular}}

\address{\normalsize $^{1}$ Institute of Imaging and Computer Vision, RWTH Aachen University, Aachen, Germany \\
\normalsize$^{2}$ Bio-manufacturing Engineering Laboratory of Tsinghua SIGS, Tsinghua University, Shenzhen, China\\
\normalsize$^{3}$ Shenzhen Eye Hospital, Shenzhen, China \\
\normalsize$^{4}$ Department of Ophthalmology, RWTH Aachen University, Aachen, Germany \\
\normalsize$^{\star}$ Equal contribution. E-mail: \href{mailto:yuli.wu@lfb.rwth-aachen.de}{yuli.wu@lfb.rwth-aachen.de} \vspace{-0.3cm}}

\begin{document}

\maketitle
\begin{abstract}

Modern biomedical image analysis using deep learning often encounters the challenge of limited annotated data. 
To overcome this issue, deep generative models can be employed to synthesize realistic biomedical images.
In this regard, we propose an image synthesis method that utilizes denoising diffusion probabilistic models (DDPMs) to automatically generate retinal optical coherence tomography (OCT) images. 
By providing rough layer sketches, the trained DDPMs can generate realistic circumpapillary OCT images. 
We further find that more accurate pseudo labels can be obtained through knowledge adaptation, which greatly benefits the segmentation task. 
Through this, we observe a consistent improvement in layer segmentation accuracy, which is validated using various neural networks.
Furthermore, we have discovered that a layer segmentation model trained solely with synthesized images can achieve comparable results to a model trained exclusively with real images. These findings demonstrate the promising potential of DDPMs in reducing the need for manual annotations of retinal OCT images.
\end{abstract}


\section{Introduction}
In practice, the non-invasive retinal optical coherence tomography (OCT) plays an important role in diagnosing eye diseases. 
Among the various features that can be extracted from OCT images, the shapes and thicknesses of multiple layers of the retina are strongly correlated with certain diseases, such as glaucoma, age-related macular degeneration and diabetic macular edema. 
Therefore, an automatic algorithm for OCT layer segmentation is of great interest, which can be achieved by deep learning~\cite{he2021retinal}.
Due to the limited availability of annotated data, which is often the case in the applications of supervised deep learning technologies, we aim to investigate the use of generative models to synthesize additional OCT images along with their corresponding ground-truth labels. 

Generative adversarial networks (GANs)~\cite{goodfellow2014generative} have demonstrated remarkable efficacy in various generative tasks, successfully replicating complex real-world content including OCT images. Many previous works leverage GANs to generate OCT images with various retinal disorders for educational or clinical purposes \cite{xiao2020open,zheng2020assessment}.
Different from GANs, flow-based deep generative models~\cite{rezende2015variational} explicitly estimate the data distribution with a sequence of invertible transformations.
Recently, denoising diffusion probabilistic models (DDPMs) \cite{ho2020denoising} have shown their great potential in generating realistic image data for various applications.
Specifically, the applicability of this concept has been demonstrated in previous work, where rough sketches of single cells or cell groups were used as a basis to generate corresponding realistic image data~\cite{eschweiler23denoising}, thereby allowing for the automated generation of fully-annotated image datasets. To this end, we can find only one study using DDPMs on OCT images from Hu \textit{et al.}\cite{hu2022unsupervised} for a plausible application of denoising.
Besides, Dhariwal \& Nichol~\cite{dhariwal2021diffusion} show that DDPMs can achieve superior sample quality compared to state-of-the-art GANs.
We thus intend to employ DDPMs to synthesize OCT images in this study.
By providing initial retinal layer sketches as input, the DDPMs trained on the annotation-free real images can produce realistic circumpapillary OCT images.
Based on visual assessment and segmentation results, however, we have identified that misregistration could occur between the layer labels from the initial sketches and the histological structures in the synthesized images. This can be explained by the inherent feature of OCT images, where the layers are adjacent, which is different from microscopy image data.
We then turn to the knowledge adaptation \cite{hu2022teacher}
to distill more accurate pseudo labels, leading to significant improvements in layer segmentation performance.
Validated using various neural networks, the layer segmentation Dice score greatly increases if we append synthetic OCT images to the real ones. Moreover, our research suggests that a layer segmentation model exclusively trained on synthesized OCT images can perform on par with a model exclusively trained on real images. These results underscore the capacity of DDPMs to enhance datasets of OCT images.

\begin{figure*}[t]
    \centering
  \includegraphics[width=0.99\textwidth]{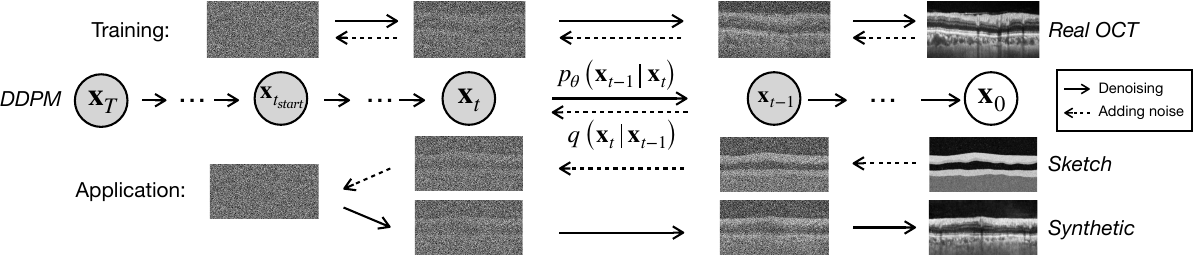}
  \caption{Generation Pipeline. The upper workflow illustrates the denoising diffusion probabilistic model (DDPM)~\cite{ho2020denoising} training process using the real retinal circumpapillary OCT images. The lower pipeline illustrates the data synthesis from the sketches with the trained DDPM. The dashed arrows from right to left denote forward diffusion processes (adding noise) and the solid arrows from left to right denote reverse diffusion processes (denoising). }
  \label{fig:pipeline}
\end{figure*}

\newpage

\section{Materials and Methods}
\subsection{Dataset}\label{sec:dataset}
We utilized the training set from the GOALS Challenge in MICCAI 2022~\cite{fang2022dataset}, which consists of 100 circumpapillary OCT images with a resolution of 1100$\times$800 pixels. 
Three layers were annotated: the retinal nerve fiber layer (RNFL), the ganglion cell-inner plexiform layer (GCIPL), and the choroid layer~(CL). The first 50 OCT images, denoted as \texttt{train-50}, are employed in this paper as the training set, while the last 50 images, denoted as \texttt{test-50}, are treated as the test set for evaluation.
Since the images cover a large section of the background, the regions of interest containing retinal layers were cropped with a fixed column height to 1100$\times$256 pixels before being downsampled to 480$\times$128 pixels, which we utilized as the input and output of all networks. The Dice scores of three retinal layers are reported during the evaluation. As instructed in the challenge, the total Dice scores of RNFL, GCIPL and CL are weighted with factors of 0.4, 0.3 and 0.3, respectively.

\subsection{Denoising Diffusion Probabilistic Models}\label{sec:medddpm}

Given the successful application of DDPMs \cite{ho2020denoising} in prior work and their capability to generate realistic image data, there is motivation for utilizing this concept in the context of this work. 
DDPMs rely on a forward diffusion process, which progressively transforms a real image data 
into pure noise by adding small noise increments at each timestep of the process:
\begin{equation}
    q(\mathbf{x}_t|\mathbf{x}_{t-1}) = \mathcal{N}\left(\mathbf{x}_t; \sqrt{1-\beta_t}\mathbf{x}_{t-1}, \beta_t\mathbf{I}\right).
\end{equation}
This process is characterized by specifying the total number of timesteps 
$T$ and the noise schedule defining the noise additions at each timestep ${\beta_t \in (0,1)}^T_{t=1}$.
By formulating this as a Markov chain with a sufficiently large number of timesteps, it becomes possible to predict the corresponding reverse diffusion process, 
thereby enabling the generation of real image data from pure noise inputs, \textit{i.e.} $\mathbf{x}_T \sim \mathcal{N}\left(\mathbf{0},\mathbf{I} \right)$. The reverse diffusion process $p_\theta$ is learned  through a neural network 
to approximate the conditional probability distributions: 
\begin{equation}
    p_\theta(\mathbf{x}_{t-1}|\mathbf{x}_{t}) = \mathcal{N}\left(\mathbf{x}_{t-1}; \boldsymbol{\mu}_\theta(\mathbf{x}_{t},t),\boldsymbol{\Sigma}_\theta(\mathbf{x}_{t},t)\right).
\end{equation}
Since the generation of image data from pure noise does not provide adequate control over the generated structures, the image generation can be adapted by short-cutting the reverse diffusion process and starting at an earlier timestep $t_{start}<T$~\cite{eschweiler23denoising}.
Optimizing this parameter allows to preserve the structures that were provided prior to the forward diffusion process and simultaneously introduce sufficient noise for a realistic texture generation throughout the reverse diffusion process.
This has the added benefit of being able to use a slightly abstract source domain, represented by sketches showing rough indications of structures and desired textural features.
Considering that the data distribution at $t_{start}$ matches the distribution of real image data at the same timestep and acknowledging that the neural network was exclusively trained on real image data, the abstract source domain gets transformed into the realistic image domain.
Consequently, the means to generate fully-annotated image datasets can be provided automatically.
The generation pipeline is demonstrated in Fig.~\ref{fig:pipeline}.

\subsection{OCT Sketch Parametrization}\label{sec:sketch}
We term a rough structure of a retinal OCT image as a \textit{sketch}, which is the input of a trained DDPM to synthesize realistic OCT images (Fig.~\ref{fig:pipeline}). Similar to the cell synthesis introduced in~\cite{eschweiler23denoising}, we initialize OCT sketches with a fixed intensity for each retinal layer. In the following, we elaborate on how the retinal OCT sketches are parameterized \textit{w.r.t.} the layer thickness, the layer intensity and two preprocessing steps: blurring and perturbation.

\noindent\textbf{Layer thickness generation}. We collect the layer segmentation ground-truth labels of \texttt{train-50} (Section~\ref{sec:dataset}) to fit a Gaussian distribution with a mean and a standard deviation of the vertical boundary coordinates for each layer. Several boundary sample pixels are randomly generated from this distribution and then smoothly connected using spline interpolation with the histological topology guaranteed.

\noindent\textbf{Layer intensity generation.}
The average intensity of each OCT image from \texttt{train-50} is chosen for each retinal layer in a sketch. We also attempted to sample from the Gaussian distribution, which yielded negligible differences.

\noindent\textbf{Blurring and perturbation.}
Two preprocessing steps are applied to the sketch images. As introduced in \cite{eschweiler23denoising}, adding Gaussian blur to sketch images is beneficial to achieve a similar data distribution earlier by smoothing unnaturally sharp boundaries. Additionally, we find that perturbing the pixel intensity from the normal distribution of each retinal layer can mimic the intrinsic noisy appearance of the OCT images.

\begin{figure}[t]
    \centering
  \includegraphics[width=0.48\textwidth]{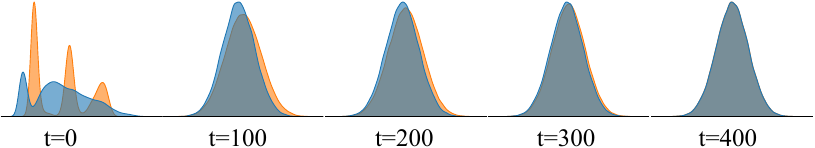}
  \caption{Histograms of a real image (blue) and a sketch (orange) \textit{w.r.t.} the timestep $t$ in the forward diffusion process. }
  \label{fig:hist}
\end{figure}

\section{Experiments And Results}
\subsection{Layer Segmentation Models}\label{sec:seg_model}
We evaluate the quality of the synthetic retinal OCT images in an end-to-end manner with multiple state-of-the-art segmentation models: U-Net~\cite{ronneberger2015u}, U$^2$-Net~\cite{qin2020u2}, FCN-ResNet~\cite{long2015fully,he2016deep}, DeepLabv3+~\cite{chen2018encoder}, and TransUNet~\cite{chen2021transunet}.
All experiments are conducted with the cross-entropy loss function.
If not otherwise specified, we generate OCT images based on \texttt{train-50} and train segmentation models with the synthesized images and labels in a supervised manner. These models are then evaluated on \texttt{test-50} with total Dice.

\subsection{DDPM Tuning}\label{sec:ddpm_tuning}
\begin{figure}[t]
    \centering
  \includegraphics[width=0.48\textwidth]{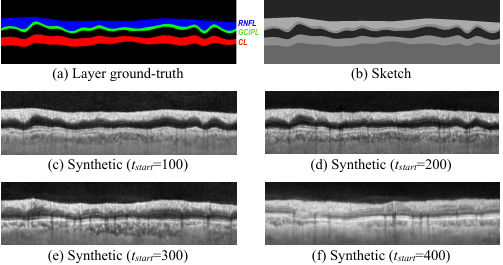}
  \caption{A layer ground-truth label image is shown in (a). The corresponding sketch image is shown in (b). Synthetic OCT images generated using this sketch image are illustrated in (c-f) based on DDPMs starting with different timesteps $t_{start}$.}
  \label{fig:sketch2t}
\end{figure}

\begin{figure}[b]
\begin{minipage}[b]{1\linewidth}
  \begin{minipage}[b]{0.59\textwidth}
    \centering
    \captionof{table}{Ablation of \textbf{B}lurring and \textbf{P}erturbation with $t_{start}$=300. The average and the best total Dice scores among 5 networks are listed.}
    \label{tab:blur}
    \begin{tabular}{cccl}
    \toprule
      \multirow{2}{0.6em}[-0.2em]{B}   & \multirow{2}{0.7em}[-0.2em]{P} & \multicolumn{2}{c}{Total Dice (in \%)}\\\cmidrule{3-4}
      && Ave.& Best \textit{(Network)} \\\midrule
        \ding{55} & \ding{55} & 72.88  &  77.13 \scriptsize{\textit{DeepLabv3+}}  \\
       \ding{51}  & \ding{55} & 72.40 &  74.88 \scriptsize{\textit{U$^2$-Net}} \\
        \ding{55} & \ding{51} & 73.40  & 77.49 \scriptsize{\textit{DeepLabv3+}}  \\
       \ding{51}  & \ding{51} & \textbf{74.65} &  \textbf{78.51} \scriptsize{\textit{U$^2$-Net}} \\
    \bottomrule
    \end{tabular}
\end{minipage}
\hfill
\begin{minipage}[t]{0.37\textwidth}
\centering
    \includegraphics[width=\textwidth]{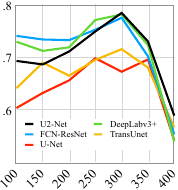}
    \captionof{figure}{Total Dice ($y$) to $t_{start}$ ($x$) with 5 networks using B and P.}
    \label{fig:tstart}
\end{minipage}
\end{minipage}
\end{figure}

We first tune the parameters of the DDPMs
\textit{w.r.t.} the starting timestep $t_{start}$, the OCT sketch parametrization and the segmentation models. We use a cosine-based scheduling of the variance $\beta_t$ \cite{nichol2021improved} and a total number of timesteps of $T=400$. A comparison of histograms in the forward diffusion process of a real image and a sketch \textit{w.r.t.} $t$ is shown in Fig.~\ref{fig:hist}, where they can hardly be distinguished at $t=400$.
With this in mind, we generate 200 synthetic OCT images from DDPMs with different timesteps $t_{start}$ of \{100, 150, 200, 250, 300, 350, 400\}. The histological layer structures of the generated OCT images are increasingly varying with an ascending $t_{start}$, as showcased in Fig.~\ref{fig:sketch2t}. Based on the visual assessment, we observe that $t_{start}=300$ is advantageous. 
An ablation study of the preprocessing steps on the sketch, namely blurring and perturbation, is conducted and listed in Tab.~\ref{tab:blur} with $t_{start}=300$ and 5 different networks. 
We find that combining both preprocessing steps yields the best total Dice. In Fig~\ref{fig:tstart}, it is further quantitatively confirmed that $t_{start}=300$ is preferable using both blurring and perturbation, which is set as the default in the following experiments.

\subsection{Distilled Pseudo Labels for OCT Synthetics}\label{sec:distill}
To this end, we can build up a fully-annotated dataset with the layer ground-truth labels from the sketches and the corresponding synthesized OCT images. 
However, we find that the layer labels do not completely colocalize with the layer structures of the synthesized OCT images from the visual assessment (Fig.~\ref{fig:sketch2t}). 
The choroid layer, shown in red in Fig.~\ref{fig:sketch2t}(a), is particularly vulnerable, as its lower boundary is not prominently visible.
It can be confirmed by the preliminary layer segmentation results trained only with the synthesized OCT images, even with an optimized $t_{start}=300$. 
Inspired by knowledge distillation \cite{hu2022teacher}, we select the best performing U$^2$-Net as a teacher model, which is pre-trained with \texttt{train-50}, to predict the layer segmentation labels for the synthesized OCT images. These predictions are then treated as the pseudo labels for the other four networks (student models). The evaluation results on \texttt{test-50} are reported in Fig.~\ref{fig:distill}, where a clear and consistent improvement can be observed between the synthetic labels from the distilled pseudo ground-truths through knowledge adaptation and the sketches. Since the teacher model is pre-trained with the real data, this approach is categorized as semi-supervised, even if we solely employ synthetic data, as shown in Fig.~\ref{fig:distill}(b). 

\subsection{Real-Synthetic Ratio}
We further experiment with different real-synthetic ratios: 50 real images to \{0, 50, 100, 200, 500, 1000\} synthesized ones (denoted as $\nicefrac{\#Real}{\#Synthetic}$). 
As listed in Tab.~\ref{tab:res}, all five models trained with a mixed dataset with 50 more synthesized OCT images ($\nicefrac{50}{50}$) surpass those with only real data ($\nicefrac{50}{0}$). 
The boost in segmentation Dice scores can be observed across most of the retinal layers. In addition, models trained on a fully-synthesized dataset ($\nicefrac{0}{1000}$) can perform on par with the ones using a real dataset ($\nicefrac{50}{0}$), two models of which perform marginally better, namely FCN-ResNet~\cite{long2015fully,he2016deep} and DeepLabv3+~\cite{chen2018encoder}. 
While the proposed pipeline can run without any annotations, we have opted, in this paper, to utilize the layer thickness and intensity distributions from the ground-truth labels as a prior to generate OCT sketches. 
Therefore, we are hesitant to classify our approach as unsupervised.

\subsection{The More, the Better?}
\begin{figure}[t]
    \centering
  \includegraphics[width=0.47\textwidth]{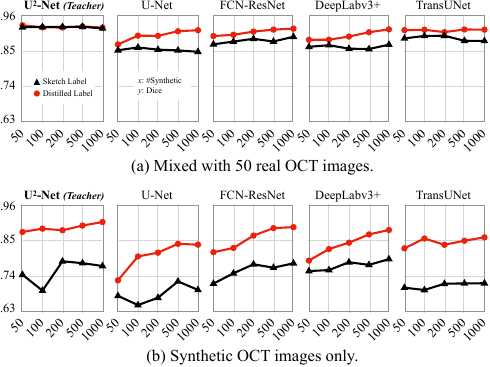}
  \caption{Total Dice scores ($y$) trained with (a) mixed synthetic and 50 real OCT images and (b) synthetic images only. The number of the synthetic images ($x$) varies from 50 to 1000.} 
  \label{fig:distill}
\end{figure}

Generally speaking, yes.
But we find that the number of the synthetic images plays a more important role, if a network is trained only with the synthesized images using distilled pseudo labels (Fig.~\ref{fig:distill}). Moreover, we argue that all sketches are parameterized based on the structural statistics from \texttt{train-50} and thus, the variety of the mixed dataset is still predominantly influenced by the real data.

\section{Discussion and Conclusion}
As addressed in Section~\ref{sec:distill}, the structures of the synthetic OCT images tend to differ from the corresponding initial sketches. Although this issue can be circumvented with knowledge adaptation via a teacher-student distillation architecture \cite{hu2022teacher} to predict more accurate pseudo labels for retinal layers, it is worthwhile to further investigate how to balance generative variety and histological structure invariance. 
Furthermore, it is promising to use it in more retinal OCT-related applications, \textit{e.g.} unsupervised domain adaptation between different OCT scanners, which was previously done with GANs \cite{wang2020domain} or contrastive learning \cite{gomariz2022unsupervised}. We can also facilitate pathological intervention in DDPMs given a retinal disorder.

Conclusively, we present an image synthesis approach that utilizes DDPMs to automatically generate realistic circumpapillary OCT images by providing initial rough layer sketches as input.
Incorporating synthetic retinal OCT images with real ones has been found to consistently improve Dice scores for layer segmentation, as confirmed by various off-the-shelf neural networks.
Our research also indicates that a layer segmentation model trained solely on synthesized OCT images can perform on par with one trained solely on real images. These findings emphasize the potential of DDPMs in reducing the reliance on manual annotations and enriching annotated datasets of retinal OCT images among various biomedical image modalities.

\begin{table}[t]
\footnotesize
\centering
\caption{Layer segmentation results with different real-synthetic ratios. The pseudo labels of the OCT synthetics are distilled through knowledge adaptation. Improved Dice scores compared to the first block ($\nicefrac{50}{0}$) are marked in bold.}
\label{tab:res}
\begin{tabular}{clrrrr}
\toprule
\multirow{2}{5em}[-0.2em]{\large$\frac{\text{\#Real}}{\text{\#Synthetic}}$} & \multirow{2}{5em}[-0.2em]{\small Network}  & \multicolumn{4}{c}{Dice (in \%)}  \\\cmidrule{3-6}
&& RNFL & GCIPL & CL & Total  \\ \midrule
\multirow{5}{4em}{\large$\nicefrac{50}{0}$} & U-Net & 90.20 &  69.80  &  92.26 & 84.70 \\
& U$^2$-Net              & 94.29 & 87.32  &    95.60 &  92.59    \\  
& FCN-ResNet     & 91.28 &  78.74  & 94.26 & 88.41 \\ 
& DeepLabv3+         & 91.36 &  76.66 & 91.11 & 86.88 \\ 
& TransUNet        & 92.35 & 84.61  &  92.58 & 90.10  \\  \midrule
\multirow{5}{4em}{\large$\nicefrac{50}{50}$} & U-Net            &  89.91 &  \textbf{77.52}  &    92.24 &  \textbf{86.89}  \\  
& U$^2$-Net              & \textbf{94.39} & \textbf{87.67}  &    \textbf{95.64} &  \textbf{92.75}    \\  
& FCN-ResNet  & \textbf{92.21} &  \textbf{80.82}  & \textbf{94.40} & \textbf{89.45} \\ 
& DeepLabv3+        & \textbf{92.12} &  \textbf{78.58} & \textbf{92.93} & \textbf{88.30} \\ 
& TransUNet      & \textbf{93.80} & \textbf{85.73}  &  \textbf{93.45} &\textbf{ 91.27} \\  \midrule
\multirow{5}{4em}{\large$\nicefrac{50}{1000}$} & U-Net            &  \textbf{93.10} &  \textbf{85.87}  &    \textbf{94.15} &  \textbf{91.25}  \\  
& U$^2$-Net              & 93.94 & 86.34  &    95.25 &  92.05    \\  
& FCN-ResNet  & \textbf{93.68} &  \textbf{85.49}  & \textbf{95.37} & \textbf{91.73} \\ 
& DeepLabv3+        & \textbf{93.49} &  \textbf{85.00} & \textbf{95.36} & \textbf{91.50} \\ 
& TransUNet      & \textbf{93.98} & \textbf{85.94}  &  \textbf{93.46} &\textbf{91.41} \\  \midrule
\multirow{5}{4em}{\large$\nicefrac{0}{1000}$} & U-Net           &  88.32 & \textbf{74.85}  &   86.09 & 83.61    \\  
& U$^2$-Net              & 91.81 & 84.50  &    94.96 & 90.56   \\  
& FCN-ResNet   & \textbf{91.71} & \textbf{81.10} & 93.28 & \textbf{89.00} \\ 
& DeepLabv3+       & \textbf{91.89} &  \textbf{76.76} & \textbf{94.46} & \textbf{88.12} \\ 
& TransUNet      & 90.96 & 75.82  &  89.11 & 85.86 \\
\bottomrule
\end{tabular}
\end{table}

\newpage
\section{Compliance with Ethical Standards}
This research study was conducted retrospectively using human subject data made available in open access by GOALS Challenge~\cite{fang2022dataset}. Ethical approval was not required as confirmed by the license attached with the open access data.

\section{Acknowledgements}
This work was supported by Deutsche Forschungsgemeinschaft (DFG, German Research Foundation) with the grant GRK2610: InnoRetVision (project number 424556709) and the grant STE2802/2-1 (project number 447699143).

\bibliographystyle{IEEEbib}
\bibliography{bib}

\end{document}